\documentstyle[aps,floats,preprint,epsfig]{revtex}

\tightenlines

\begin{document}
\draft
\preprint{\vbox{\hbox{TRI-PP-03-02}
\hbox{January 2003}}}
\title{Resolved Photon Contributions to Higgs Boson Production\\ 
in $\gamma\gamma$ Collisions}
 \author{M. A. Doncheski}
\address{Department of Physics, Pennsylvania State University, \\
Mont Alto, PA 17237 USA}
\author{Stephen Godfrey}
\address{
Ottawa-Carleton Institute for Physics \\
Department of Physics, Carleton University, Ottawa, Canada K1S 5B6 \\
{\it and} \\
TRIUMF, 4004 Wesbrook Mall, Vancouver B.C. Canada V6T 2A3}

\maketitle

\begin{abstract}
We study single Higgs boson production in $\gamma\gamma$ collisions proceeding via 
the hadronic content of the photon.  These processes complement previous 
studies of pair and single Higgs production at $e^+e^-$ colliders.  For SM 
Higgs masses of current theoretical interest, the resolved photon 
contributions are non-negligible in precision cross section measurements.  
The charged Higgs cross sections are not competitive with the 
$\gamma\gamma \to H^+ H^-$ process and at best might offer some information 
about quark-Higgs couplings.  Finally, resolved photon production of the 
heavier Higgs bosons, $H^0$ and $A^0$ of the MSSM, can probe regions of the 
SUSY parameter space that will complement other measurements.  This 
last process 
shows some promise in SUSY Higgs searches and warrants further study.
\end{abstract}
\pacs{PACS numbers: 12.15.Ji, 12.60.Cn, 14.70.-e, 14.80.-j}

\section{Introduction}

One of the fundamental questions of the standard model (SM) of particle 
physics is the origin of electroweak symmetry breaking 
(EWSB)\cite{dawson99,carena02}.  
The simplest description of EWSB results in one neutral scalar particle, the Higgs boson, which 
has well known problems associated with it.  A priori, a more 
complicated Higgs sector is phenomenologically just as viable.  The next 
simplest case is the general two Higgs doublet model (2HDM).  A constrained 
version of the 2HDM arises in the minimal supersymmetric extension of the 
SM (MSSM) \cite{susy} where spontaneous symmetry breaking is induced by two 
complex Higgs doublets and leads to five physical scalars; the neutral CP-even 
$h^0$ and $H^0$ bosons, the neutral CP-odd $A^0$ boson, and the charged 
$H^\pm$ bosons.  At tree level the MSSM Higgs sector has two free parameters 
which are usually taken to be the ratio  of the vacuum expectation values of 
the two Higgs doublets, $\tan\beta=v_2/v_1$, and the mass of the $A^0$ boson, 
$m_A$.  The elucidation of  EWSB is the primary goal of the Large Hadron 
Collider at CERN (LHC). 

Direct searches at LEP2 yield lower limits $M_H> 114.4$~GeV for the SM Higgs 
mass \cite{lep2} and $M_{h,A} \gtrsim 90$~GeV for the neutral SUSY Higgs masses while 
$M_{H^\pm} \gtrsim 80-90$~GeV depending on the $H^\pm$ decay modes.  More detailed 
results have been presented by the LEP Collaborations as plots of excluded 
regions of the MSSM parameter space \cite{lep2}.  Similar plots have been 
obtained by the CDF and D0 collaborations at the Tevatron $p\bar{p}$ collider 
\cite{tevatron}.  It is expected that  Run II of the Tevatron $p\bar{p}$ 
collider will be able to find evidence at $3\sigma$ for the SM Higgs boson 
up to about 180~GeV  although a $5\sigma$ signal 
 is limited to around 130~GeV \cite{run2}.

In proton-proton collisions at $\sqrt{s}=14$~TeV at the LHC the ATLAS 
\cite{atlas} and CMS \cite{cms} experiments have shown that they are sensitive 
to the SM Higgs boson over the entire mass range of 100-1000~GeV.  The MSSM 
Higgs boson can be discovered in a variety of channels so that at least one 
Higgs boson can be discovered for the entire parameter range.  In a fraction 
of the parameter space more than one Higgs boson is accessible.  However, there 
is a region in which the extended nature of the supersymmetric Higgs sector 
may not be observable since only the lightest Higgs boson can be seen 
in SM-like 
production processes.  Likewise, only a limited number of measurements of 
Higgs boson properties can be carried out at the LHC.

A future high energy linear $e^+e^-$ collider has been proposed as an 
instrument that can perform precision measurements that would complement those 
performed at the LHC \cite{teslatdr,higgs}.  In this context, the 
photon-photon ``Compton-collider'' option, from backscattered laser light off 
of highly energetic and possibly polarized electron beams, has been advocated 
as a valuable part of the LC physics program \cite{telnov}.  Recently, a 
number of papers have shown how the Compton-collider option can make important 
measurements in the Higgs sector 
\cite{ggtoh,ggtomssm,melles,guo99,asner,krawczyk,fei01}.  
For example, the analysis by M\"uhlleitner, {\it et al.}, 
shows that the $\gamma-\gamma$ option of TESLA can be used to produce the 
$h^0$, $H^0$, and $A^0$ for intermediate values of $\tan\beta$ that may escape 
discovery at the LHC \cite{ggtomssm,ggtoh}.

In this paper we show that the hadronic content of the photon can result in 
cross sections large enough that they should be considered in precision 
measurements of $\gamma\gamma\to H$ cross sections and that they may be useful 
in the study of the Higgs sector of the theory.  The importance of resolved 
photon contributions has been demonstrated by the OPAL \cite{lqopal} and 
DELPHI \cite{lqdelphi} collaborations in obtaining interesting limits on 
leptoquark properties from $e\gamma $ production of leptoquarks \cite{lq}.

\section{Calculations and Results}

The resolved photon approach is the same for all processes we  
consider so we begin with a brief description of the generic approach shown 
in Fig. 1.  In the resolved photon approach the quark and gluon content of the 
photon are treated as partons described by partonic distributions, 
$f_{q/\gamma}(x,Q^2)$ in direct analogy to partons inside hadrons \cite{fph}. 
The parton subprocess cross sections
are convoluted with the parton distributions to 
obtain the final cross sections.  However, since the photons themselves have 
an associated spectrum, either the energy distribution obtained from 
backscattering a laser from an electron beam \cite{backlaser} or the 
Weizs\"{a}cker Williams distribution \cite{WW}, we must further convolute the 
cross sections with the photon distributions to obtain cross sections that can 
be compared to experiment.  Thus, the cross sections are found by evaluating 
the following expression:
\begin{equation}
\label{cross}
\sigma = \int dx_1 dx_2 dx_3 dx_4 f_{\gamma/e}(x_1,Q^2)
f_{\gamma/e}(x_2,Q^2) f_{p_i/\gamma}(x_3,Q^2)
f_{p_j/\gamma}(x_4,Q^2) \; \hat{\sigma}(\hat{s})
\end{equation}
where $p_{i(j)}$ represents parton $i(j)$ in the photon which could be a 
quark, anti-quark, or gluon, $\hat{\sigma}(\hat{s})$ represents the subprocess 
cross section with parton C of M energy $\sqrt{\hat{s}}$.

The processes we are studying are typically of the form $q\bar{q}\to h^0$ and 
therefore take a particularly simple form.  In this preliminary study we only 
include tree-level contributions and are aware that higher order
corrections are 
likely to be non-negligible.  Nevertheless, we feel that our approach is 
satisfactory for a first survey of resolved photon contributions to identify 
which processes may warrant more detailed study.  We will adopt the standard 
used in many  LC studies that 20 events represents an interesting signal for 
the canonical integrated luminosities used in such studies of 1~ab$^{-1}$.  We 
do not include branching ratios, detector efficiencies, nor consider 
backgrounds.  In the former case the branching ratios are well known 
\cite{higgsbr,mssmdecays} and we assume that the environment is sufficiently 
clean that the signal can be reconstructed with reasonable efficiency.  
Clearly, our study is crude and a more detailed study is warranted which 
includes a realistic consideration of detector acceptance 
and efficiency assumptions, decays 
leading to final state particles, and background studies.

Our calculations have explicit dependence on the $c$ and $b$-quark masses.  We 
take $m_s=0.15$~GeV,
$m_c=1.4$~GeV, $m_b=4.4$~GeV, $V_{cs}=0.97$,
and $V_{bc}=0.04$ \cite{pdg}.  In addition, we used 
$M_W=80.41$~GeV,  $G_F=1.166\times 10^{-5}$~GeV$^{-2}$, $\alpha=1.0/128.0$, and 
$m_t=175.0$~GeV.

\subsection{Standard Model Higgs Production}

We begin with SM Higgs production because the discovery and elucidation of the 
standard model Higgs boson is the first order of business for any future 
collider project.  One of the strongest motivations for the Compton collider 
is to measure Higgs boson properties.  The $2\gamma$ production of the Higgs 
boson is an especially interesting reaction \cite{ggtoh,ggtomssm,melles} as it 
proceeds via loop contributions \cite{loops} and is therefore sensitive to new 
particles much higher in mass than the CofM energy that cannot be 
produced directly.  It has also been
suggested that this reaction can be used to 
distinguish between the SM Higgs and the lightest scalar Higgs of the MSSM in 
the decoupling limit in which no other Higgs or SUSY particles are observed at 
the LC \cite{decoupling}.  It is therefore
important that all SM contributions to this cross section be 
carefully considered.

The first process we consider is $q\bar{q} \to H$ where the quark 
and anti-quark arise from the quark parton distributions of the photon, the so 
called resolved photon processes.  The expression for the subprocess cross 
section is rather trivial and is given by:
\begin{equation}
\label{qqtoh}
\sigma(q\bar{q}\to H)  =  \frac{G_F \pi}{3\sqrt{2}} m_f^2
\sqrt{(1-4m_f^2/M_h^2)} \;
\delta(M_H^2-\hat{s}) 
\end{equation}
The cross section is dominated by the $c$-quark content.  The lower mass of 
the $c$-quark enters eqn. \ref{qqtoh} quadratically, but its charge and mass 
enhance the  $c$-quark content in the photon.  In Fig. 2 we show, for the 
backscattered laser case with $\sqrt{s_{ee}}=500$~GeV, 1~TeV and 1.5~TeV using 
the GRV distribution functions \cite{grv}, the contributions from  both 
$c\bar{c}$ and $b\bar{b}$ production and the sum of the two.  
We also show the cross sections for the subprocess $\gamma\gamma\to H 
t\bar{t}$ \cite{cheung93} which we 
calculated using the COMPHEP computer package 
\cite{comphep}.  The contributions from lighter quarks (including the $b$) are 
dominated by the collinear region of $\gamma\gamma\to H q\bar{q}$, and as such 
are well described via the resolved photon approach.  
On the other hand, due to its large mass, 
the $t$-quark avoids the collinear region so that the 
resolved photon approach is inappropriate and we use the full subprocess.
For the Higgs masses we consider, this 
process is below threshold for $\sqrt{s}_{ee}=500$~GeV due to the 
t-quarks produced. For $\sqrt{s}_{ee}=1000$~GeV, where there is more 
available phase space, this mechanism is 
interesting for Higgs masses up to about 200~GeV  and up to about 
300~GeV for $\sqrt{s}_{ee}=1500$~GeV.  Whether these two cases are 
interesting experimentally will depend crucially on the t-quark 
tagging efficiency.

For comparison 
we also show the $\sigma(\gamma\gamma\to H)$ production which proceeds via 
loops.  The expressions are well known in the literature \cite{HHguide,barger} 
and we do not reproduce them here.  Likewise, we show the contribution from 
gluon fusion, $\sigma(gg\to H)$, which arises from the gluon content of the 
photon.  Again, the subprocess cross section is well known in the literature.
A final process that will produce Higgs bosons is 
$\gamma\gamma\to H W^+W^-$ \cite{cheung94,jikia95}. The cross section 
was calculated using  COMPHEP \cite{comphep} and is shown as the 
dot-dashed curves in Fig. 2.  The cross sections are seen to be 
substantial, rivalling the dominant $\gamma\gamma\to H$ at 
$\sqrt{s}_{ee}=1500$~GeV although it is much smaller at
$\sqrt{s}_{ee}=500$~GeV, where it is comparable to the resolved photon 
processes we are interested in.
To some extent this process can be disentangled from the 
$\gamma\gamma \to H$ process and the resolved photon processes via $W$ 
and jet tagging.  Nevertheless it is yet another ingredient that 
should be considered in measuring the Higgs two-photon width.

The two photon process dominates the resolved photon processes
over the full range of Higgs masses we 
consider.  The gluon-fusion process is almost three orders of magnitude 
smaller than  $\sigma(\gamma\gamma \to H)$ over the entire $M_H$ range.  We 
note that for $\sqrt{s}=1500$~GeV the gluon-fusion contribution increases to 
about 10\% of the total $q\bar{q}$ contributions  for $M_H < 400$~GeV.  This 
reflects the increasing importance of the small $x$ contributions from the 
gluon distributions at higher $\sqrt{s}$.  Nevertheless, it does not seem 
unreasonable to neglect the gluon fusion contributions in our results.  
In contrast, the quark annihilation processes contribute at the percent level 
for $M_H\sim 150$~GeV and $\sqrt{s_{ee}}=500$~GeV.  This increases to the 
several percent level for $\sqrt{s_{ee}}=1.5$~TeV.  This Higgs mass region has 
received considerable attention for study at a future Linear Collider and 
Compton Collider and  measurements of the $\gamma\gamma \to H$ cross section 
at this level of precision is touted as a probe of new physics entering via 
loops.  Recent studies by Asner {\it et al} \cite{asner} 
and by Krawczyk {\it et al} \cite{krawczyk} find that these processes 
can be measured to approximately the  2\% level.
Thus, depending on $M_H$, Higgs production via the hadronic 
content of the photon may not be
negligible at this level of precision, suggesting that these
contributions deserve further study.

There exist many different sets of photon parton distributions in the 
literature
\cite{fph,do,dg,lac,resphot}.  In Fig. 3 we give an indication of the 
uncertainties due to the distribution functions by plotting the variation in 
cross section using different photon parton distribution functions.  
The solid band 
and shaded bands are for $\sigma(c\bar{c}\to H)$, $\sigma(b\bar{b}\to H)$, and 
$\sigma(gg\to H)$ using the DO \cite{do}, GRV \cite{grv}, DG \cite{dg}, LAC1 
and LAC2\cite{lac} distributions.  This, by no means represents a complete 
survey of available distributions and is only meant to illustrate the 
uncertainties inherent in our current knowledge.  It can be seen that there is 
considerable variation in the cross section depending on the specific 
distribution used.  Fortunately, with significant amounts of new data from 
the LEP and HERA experiments, revised distributions based on these 
new data are arriving \cite{cornet02}.
To take a conservative approach we used
the GRV distributions which generally give the smallest cross sections.

In addition to the Compton Collider configuration, $\gamma\gamma$ luminosities 
arise at  $e\gamma$ and $e^+e^-$ colliders.  In $e\gamma$ collisions 
one $\gamma$ comes from a backscattered laser as before, while the second 
photon is a Weizs\"acker-Williams bremsstrahlung photon.  In $e^+e^-$ both 
photons are Weizs\"acker-Williams photons. The Weizs\"acker-Williams 
photon spectrum is
softer than the backscattered laser spectrum but it extends to higher 
$x=E_\gamma/E_e$.  In Fig. 4 we show the SM Higgs production cross section as 
a function of $M_H$ for the 3 cases.  
As we go from $\gamma\gamma$ to $e\gamma$ to $e^+e^-$ 
the cross section decreases by about an order of magnitude in each case 
although at larger values of $M_H$, near the kinematic limit where the 
$\gamma\gamma$ luminosity goes to zero, the $\gamma\gamma$ luminosity extends 
further out for the $e\gamma$ and $e^+e^-$ cases.  We find similar patterns 
for charged Higgs production and the production of the heavier Higgs bosons of 
the MSSM model.  We will therefore only show cross sections for the 
$\gamma\gamma$ case, 
although for completeness we mention results for the $e^+e^-$ and 
$e\gamma$ cases when appropriate.

\subsection{Charged Higgs Boson Production}

Charged Higgs bosons arise naturally from the simplest extension of the SM: 
the introduction of a second Higgs doublet.  Two variations of the 
two doublet model are discussed in the literature:  In Model I, $v_2$ 
couples to both the $u$ and $d$-type quarks and the other field 
decouples. 
In Model II, which arises in the MSSM, $v_2$ couples to the up-type 
quarks and $v_1$ to the down-type quarks. The ratio of these VEV's is 
one of the fundamental parameters of the theory; $\tan\beta\equiv 
v_2/v_1$.
In this paper we restrict ourselves to model II. 
A charged Higgs boson will be difficult to find at the LHC if 
its mass is greater than $\sim 125$~GeV.  This discovery reach is extended  
to almost $M_{H^\pm} \sim \sqrt{s}/2$ at an $e^+e^-$ linear collider where 
they can be pair produced in the process $e^+e^- \to H^+ H^-$ with a cross 
section which depends mainly on the $H^\pm$ mass, dropping quickly near 
threshold due to $P$-wave suppression \cite{chargedhiggs,komamiya88}.  
Charged Higgs bosons can also be pair produced in the process
$\gamma\gamma\to H^+H^-$ \cite{chao93}
Given 
the interest in their properties, there have been a number of recent 
studies of single charged Higgs bosons at LC's 
\cite{kanemura,moretti02,he02}.  
If charged Higgs 
bosons were observed, cross section measurements could potentially give 
information about the underlying theory.

As before the subprocess cross section is rather straightforward to calculate 
and for $b\bar{c}$ fusion in Model II is given by:
\begin{equation}
\label{bctoch}
\sigma(b\bar{c}\to H^-)  =  {{G_F \pi}\over{3\sqrt{2}}} |V_{bc}|^2 
 \frac{(m_b^2 \tan^2\beta +m_c^2 \cot^2\beta)(M_H^2 - m_b^2 -m_c^2) - 4 m_b^2 m_c^2}
{\sqrt{(M_H^2 - m_b^2 - m_c^2)^2 - 4 m_b^2 m_c^2}} 
\; \delta(M_H^2-\hat{s}) 
\end{equation}
with an analogous expression for $s\bar{c}$ fusion.
We include in our results a factor of 2 to take into account that the 
$b$-quark can come from either photon and likewise for the $c$ anti-quark
and a 2nd factor of 2 for summing over $H^-$ and $H^+$ 
production.  
A priori one would expect the $bc$ process to dominate over the 
$cs$ process due to the much larger mass of the $b$-quark compared to 
the $s$-quark mass.  However, the mass ratios $m_s/m_b$ and $m_c/m_b$ 
are compensated 
by the ratio of the CKM matrix elements $V_{cs}/V_{cb}$ so that 
$m_s/m_b \times V_{cs}/V_{cb} \simeq 0.8$ and 
$m_c/m_b \times V_{cs}/V_{cb} \simeq 8$.  For small values of 
$\tan\beta$ the $cs$ fusion process will be larger than the $cb$ 
fusion process.
In addition to $bc$ and 
$cs$ fusion, charged Higgs bosons can also be produced via $bt$ 
fusion.  The $t$-quark is too heavy to be treated as a 
massless constituent of the photon 
so in this case we calculate the cross section for $b\gamma \to H^- t$.  
Although this last process will have a kinematic limit 
constrained by the Higgs and $t$-quark mass, below threshold it is the 
dominant contribution due to the large $t$-quark mass \cite{doncheski}.  
Above the $H-t$ threshold the relative importance of the $cs$ and $cb$ 
subprocesses is dependent on $\tan\beta$ via eqn \ref{bctoch}.

The various contributions are shown in Fig. 5 for $\tan\beta=3.0$ and 
$40$ for $\sqrt{s}_{ee}=500$~GeV.  
It should be noted that the cross sections for $\tan\beta=1.5$ 
are larger than those for $\tan\beta=3.0$.  
In all cases, for relatively low $M_H$,
 the largest contribution to the total charged 
Higgs production cross section comes from the $tb$ contribution.  As 
the Higgs mass increases the other contributions become dominant until 
the $tb$ contribution goes to zero at the $t$-$H$ kinematic limit.
For small values of $\tan\beta$ the $cs$ contribution is larger than 
the $bc$ contribution while for large values of $\tan\beta$ the $cs$ and 
$cb$ contributions are comparable in size.  In Fig. 6 we show the sum 
of these three contributions for a range of values of $\tan\beta$.  
For comparison, we also show the cross sections for $e^+e^-\to H^+ 
H^-$ (medium dashed line) \cite{komamiya88}
and for $\gamma\gamma\to H^+ H^-$ 
(dot-dashed line) \cite{blundell,chao93}.

Using the criteria adopted by the TESLA TDR of 20 events for 1~ab$^{-1}$ of 
integrated luminosity we find that for $\sqrt{s}_{ee}=500$~GeV, 
a charged Higgs bosons can be detected in this 
process up to $M_H=270$~GeV and 240~GeV for $\tan\beta=$40, and 30 
respectively.  As we have not taken into account acceptance cuts nor have we 
examined background reduction these numbers should be taken with a grain of 
salt.  A proper analysis would require detailed simulations including detector 
dependent considerations which are clearly beyond the scope of the present 
work.  With this caveat, for the high $\tan\beta$ cases, a charged Higgs boson 
could be detected with mass 
roughly comparable to the standard kinematic limit associated 
with charged particle 
pair production at $e^+e^-$ colliders.  However, these cross sections 
are greater than most of the processes considered by Kanemura, {\it et al.} 
\cite{kanemura} so that cross section information from this 
production mechanism would 
offer complementary information to other processes. For 
$\tan\beta >1$, charged Higgs boson production in Model I would be too small to 
be observed.

We also considered the case where the photons bremsstrahlunged off of the 
incident $e^+ e^-$ beams using the Weizs\"acker-Williams effective photon 
distribution.  In this case the cross sections would only produce 
measurable event rates for Higgs masses much smaller than could be 
produced in the pair production process $e^+e^-\to H^+H^-$.
The 
intermediate case, for an $e\gamma$ collider, where one photon radiates off an 
initial beam electron and the other obtained from backscattering a laser off 
of the initial electron beam results in a cross section intermediate in 
magnitude between the $\gamma\gamma$ case and $e^+e^-$ case.
Using the same 
criteria as above, for $\sqrt{s}_{ee}=500$~GeV
we obtain measurable rates for $M_H< 145$ and 165~GeV for the 
$\tan\beta=$30 and 40 cases.  
The discovery limits for various collider energies are summarized in 
Table I.
Given that $\sigma(e^+e^-\to H^+H^-)$ is significantly
larger than these cross sections it is unlikely that much would be 
learned from this process in $e\gamma$ collisions.

Motivated by the article of Kanemura, {\it et al.} \cite{kanemura},
we also calculated the resolved photon contributions to $H^\pm W^\mp$ 
production which proceeds via $q\bar{q}$ fusion through an intermediate
neutral Higgs boson.
However, the cross sections were found to be small
with event rates  uninteresting from an experimental point of view.

\subsection{Heavy MSSM Neutral Higgs Boson Production}

In two Higgs doublet models there exist a total of three neutral Higgs bosons 
in addition to the two charged Higgs bosons discussed above.  The three 
neutral Higgs bosons consist of two CP-even bosons, $h$ and $H$, the former 
light and the latter heavy, and a heavy
CP-odd boson $A$.  At the CERN LHC, the 
pseudoscalar Higgs is not detectable above $M_A \gtrsim 250$~GeV 
for intermediate 
values of $\tan\beta$ \cite{spira}.  The MSSM Higgs bosons can be produced in 
$\gamma\gamma$ collisions, $\gamma \gamma \to h^0, \; H^0, \; A^0$, with 
favorable cross sections allowing a heavy Higgs boson to be found up to 
70-80\% of the initial $e^+e^-$ collider energy for moderate values of 
$\tan\beta$ \cite{ggtomssm}.  However, it turns out that the resolved 
photon process has different dependence on $\tan\beta$ than the process 
proceeding through intermediate loops \cite{guo99b}.  
For example, the two-photon decay 
width of the $A^0$ is small for large $\tan\beta$ while the decay width of the 
$H^0$ is small for small $\tan\beta$ \cite{higgsbr}.  For the resolved photon 
processes, for large $M_A$ the $A^0$ and $H^0$ have roughly similar cross 
sections independent of $\tan\beta$.

We use the minimal supersymmetric standard model (MSSM), to partially 
constrain the parameters of the model in our calculations 
\cite{HHguide,mssm}.  Taking $\tan\beta$ and $M_A$ or $M_{H^\pm}$ as 
input, the rest of the parameters can be calculated.  For simplicity, and 
given the other uncertainties in our results, we use tree level 
relationships for the various parameters.  These are given by:
\begin{eqnarray}
\label{susy}
& & M_{H^\pm}^2 = M_A^2 + M_W^2 \nonumber\\
& & M^2_{H^0, h^0} = \frac{1}{2} [M_A^2 + M_Z^2 \pm 
\sqrt{(M_A^2 + M_Z^2)^2 - 4M_Z^2 M_A^2 \cos^2 2\beta}] \nonumber\\
&& \cos 2\alpha =-\cos 2\beta \left( { {M_A^2 -M_Z^2}\over {M_H^2 
-M_h^2}} \right)
\end{eqnarray}
The Higgs couplings  to quarks are given by:
\begin{eqnarray}
\label{couplings}
h^0 c\bar{c}: & & {{-igm_c \cos\alpha}\over {2M_W\sin\beta}} \nonumber \\
h^0 b\bar{b}: & & {{igm_b \sin\alpha}\over {2M_W\cos\beta}} \nonumber \\
H^0 c\bar{c}: & & {{-igm_c \sin\alpha}\over {2M_W\sin\beta}} \nonumber \\
H^0 b\bar{b}: & & {{-igm_b \cos\alpha}\over {2M_W\cos\beta}} \nonumber \\
A^0 c\bar{c}: & & {{-gm_c \cot\beta}\over {2M_W}} \gamma_5\nonumber \\
A^0 b\bar{b}: & & {{-gm_b \tan\beta}\over {2M_W}} \gamma_5 
\end{eqnarray}

These couplings result in the following subprocess cross sections:
\begin{eqnarray}
\label{susycs}
\sigma(c\bar{c} \to h^0) & = & {{G_F \pi}\over {3\sqrt{2}}} m_c^2 
{{\cos^2\alpha}\over {\sin^2 \beta}} \; \delta (M_h^2 -\hat{s}) \nonumber \\
\sigma(b\bar{b} \to h^0) & = & {{G_F \pi}\over {3\sqrt{2}}} m_b^2 
{{\sin^2\alpha}\over {\cos^2 \beta}} \; \delta (M_h^2 -\hat{s}) \nonumber \\
\sigma(c\bar{c} \to H^0) & = & {{G_F \pi}\over {3\sqrt{2}}} m_c^2 
{{\sin^2\alpha}\over {\sin^2 \beta}} \; \delta (M_H^2 -\hat{s}) \nonumber \\
\sigma(b\bar{b} \to H^0) & = & {{G_F \pi}\over {3\sqrt{2}}} m_b^2 
{{\cos^2\alpha}\over {\cos^2 \beta}} \; \delta (M_H^2 -\hat{s}) \nonumber \\
\sigma(c\bar{c} \to A^0) & = & {{G_F \pi}\over {3\sqrt{2}}} m_c^2 
\cot^2\beta \; \delta (M_A^2 -\hat{s}) \nonumber \\
\sigma(b\bar{b} \to A^0) & = & {{G_F \pi}\over {3\sqrt{2}}} m_b^2 
\tan^2\beta \; \delta (M_A^2 -\hat{s}) \nonumber \\
\end{eqnarray}

The cross sections for $H^0$ and $A^0$ production via $b\bar{b}$ and 
$c\bar{c}$ annihilation  for $\tan\beta =3$ and 30 are shown in Fig 7.  For 
higher values of $\tan\beta$, $b\bar{b}$ annihilation dominates, while for lower 
values, $c\bar{c}$ annihilation becomes more and more important until at 
sufficiently small values of $\tan\beta$ the $c\bar{c}$ annihilation 
contribution will become larger than the $b\bar{b}$ contribution. 
%
%
%
Fig. 8 shows the sum of the $b\bar{b}\to H$ and $c\bar{c}\to H$ contributions 
for a variety of $\tan\beta$ values for $\sqrt{s}_{ee} =500$ and 1000~GeV.  
For a $\gamma\gamma$ collider with $\sqrt{s_{ee}}=500$~GeV, again using the 
criteria of 20 events for integrated luminosity of 1~ab$^{-1}$, the $H^0$ can 
be observed up to 375~GeV, 365~GeV, 275~GeV, 175~GeV and 135~GeV for 
$\tan\beta=$ 40, 30, 7, 3, and 1.5 respectively.  The cross section is totally 
dominated by $b\bar{b}$ annihilation for the three large $\tan\beta$ cases 
while for $\tan\beta=1.5$ the $c\bar{c}$ contribution is actually larger for 
the values of $M_H$ with large enough cross section to allow for discovery.  
The relative importance crosses over for $M_H>120$~GeV but for the full range 
of $M_H$ shown, the $c\bar{c}$ contribution is not negligible.


The cross sections for $A^0$ production are very similar to the $H^0$ cross 
sections so we do not show them but simply summarize the results in 
Table II. 

For both $H$ and $A$ production, for the larger values of $\tan\beta$, the 
$H^0$ and $A^0$ will be produced in substantial numbers for masses 
substantially larger than $\sqrt{s}/2$.  Even if one uses a much more 
stringent criteria for discovery than the one we have adopted here, we expect 
that the heavy Higgs bosons can be observed with relatively
high masses.
This is in contrast to the SM 
Higgs production where the resolved photon contributions are at best a 
non-negligible contribution that need to be understood in precision 
measurements.  If $H$ and $A$ were produced in sufficient quantity it is 
possible that the cross section could be used to constrain $\tan\beta$ in 
analogy to the proposal by Barger, {\it et al.} to use heavy Higgs productions 
to determine $\tan\beta$ \cite{barger01}.  Thus, the resolved photon 
contributions may very well play an important role in understanding the Higgs 
sector.

This can be seen most clearly by showing the regions of the 
$\tan\beta - M_{A}$ plot which can be explored via $H$ and $A$ production.  
Plots are given in Fig. 9 for $\sqrt{s}_{ee}=500$ and 1000~GeV.  Three regions 
are shown.  Region 1 is for $\sigma > 0.1$~fb which would result in $>100$ 
events for 1~ab$^{-1}$, region 2 is for $\sigma > 0.02$~fb which would result 
in $>20$ events while in region 3 less than 20 events would be expected.  The 
regions covered would complement measurements made in other processes.

\section{Conclusions}

The resolved photon contributions to two photon production of the SM Higgs 
boson is non-negligible for the more probable Higgs masses found in 
electroweak fits.  Given that this process it touted as a sensitive probe of 
new physics via loop contributions it is important that the resolved photon 
contributions be understood at the same level as the loop contributions.  

The 
resolved photon production of charged Higgs is unlikely to be an important 
production mechanism.  At best, if $M_{H^\pm}$ and $\tan\beta$ take on certain 
values, it may offer complementary information to other processes.

Resolved photon production of the heavy Higgs bosons in 
the MSSM are potentially 
the most interesting processes.  They can be produced up to relatively high 
mass and in regions of MSSM parameter space that complements other 
measurements.

Our results motivate further study including decay modes, the hadronic final 
states and the backgrounds relevant to these single Higgs boson production 
processes.

\acknowledgments

The authors thank Sally Dawson, Pat Kalyniak, Maria 
Krawczyk,  Wade Hong, and Uli Nierste for useful comments and discussions.
This research was supported in part by the Natural Sciences and Engineering 
Research Council of Canada.  The work of M.A.D.\ was supported, in part, by 
the Commonwealth College of The Pennsylvania State University under a Research 
Development Grant (RDG).

\newpage
\begin{table}
\caption{The charged Higgs masses (in GeV) corresponding to 
$2\times 10^{-2}$ fb  and $10^{-1}$ fb cross sections.}
\begin{center}
\begin{tabular}{llrrrr}
$\sqrt{s}$ & $\tan\beta$  & \multicolumn{2}{c}{30} &\multicolumn{2}{c}{40}\\
\hline
	& $\sigma$ & $2\times 10^{-2}$ fb & $ 10^{-1}$ fb 
	& $2\times 10^{-2}$ fb & $ 10^{-1}$ fb \\
\hline
500 GeV & $\gamma\gamma$  &  240 & 180 & 270 & 200  \\
     	& $e\gamma$       &  145 & 80  & 165 & 105  \\
\hline
800 GeV & $\gamma\gamma$  &  410 & 345 & 425 & 375  \\
     	& $e\gamma$       &  270 & 155 & 305 & 195  \\
\hline
1000 GeV & $\gamma\gamma$  &  540 & 455 & 560 & 490  \\
     	 & $e\gamma$       &  350 & 205 & 395 & 255  \\
\hline
1500 GeV & $\gamma\gamma$  &  855 & 705 & 890 & 765  \\
     	 & $e\gamma$       &  525 & 305 & 605 & 385  \\
\end{tabular}
\end{center}
\end{table}

\newpage
\begin{table}
\caption{The
$H^0$ and $A^0$ Higgs masses (in GeV) corresponding to 
$2\times 10^{-2}$ fb  and $10^{-1}$ fb cross sections.
}
\begin{center}
\begin{tabular}{|lll|rrrrr|rrrrr|} 
	&	&	& \multicolumn{5}{c}{$M_H$} & 
				\multicolumn{5}{c}{$M_A$} \\
\hline
$\sqrt{s}$ & $\tan\beta$& $\sigma$ (fb) &  1.5 & 3 & 7 & 30 & 40 &
				 1.5 & 3 & 7	& 30 & 40 \\
\hline
500 GeV & $\gamma\gamma$ & 0.02 & 
	135 & 175 & 275 & 365 & 375 & 125 & 180 & 275 & 365 & 375 \\
	& 	 & 0.1 & 
	--- & --- & 180 & 335 & 350 & --- & 100 & 185 & 330 & 345 \\
     	& $e\gamma$   & 0.02 &
	--- & --- & 125 & 275 & 295 & --- & --- & 135 & 270 & 295 \\
	& 	 & 0.1 & 
	--- & --- & --- & 190 & 220 & --- & --- & --- & 190 & 220 \\     
     	& $e^+e^-$    	 &  0.02 & 
	--- & --- & --- & 140 & 165 & --- & --- & --- & 140 & 160 \\
	& 	 & 0.1 & 
	--- & --- & --- & --- & 100 & --- & --- & --- & --- & 100 \\
\hline
800 GeV & $\gamma\gamma$ & 0.02 & 
	140 & 220 & 375 & 560 & 580 & 140 & 220 & 380 & 560 & 580 \\
	& 	 & 0.1 & 
	--- & --- & 230 & 480 & 520 & --- & 115 & 220 & 475 & 510 \\
     	& $e\gamma$   & 0.02 &
	--- & --- & 165 & 375 & 420 & --- & 95 & 170 & 380 & 420 \\
	& 	 & 0.1 & 
	--- & --- & --- & 250 & 295 & --- & --- & 95 & 250 & 390 \\     
     	& $e^+e^-$    	 &  0.02 & 
	--- & --- & --- & 195 & 225 & --- & --- & 80 & 195 & 220 \\
	& 	 & 0.1 & 
	--- & --- & --- & 115 & 140 & --- & --- & --- & 110 & 140 \\
\hline
1000 GeV & $\gamma\gamma$ & 0.02 & 
	155 & 240 & 430 & 685 & 710 & 155 & 240 & 430 & 685 & 710 \\
	& 	 & 0.1 & 
	--- & 100 & 240 & 560 & 610 & --- & 120 & 250 & 570 & 620 \\
     	& $e\gamma$   & 0.02 &
	--- & --- & 185 & 435 & 495 & --- & 100 & 190 & 440 & 500 \\
	& 	 & 0.1 & 
	--- & --- & --- & 280 & 330 & --- & --- & 100 & 280 & 340 \\     
     	& $e^+e^-$    	 &  0.02 & 
	--- & --- & --- & 220 & 255 & --- & --- & --- & 220 & 255 \\
	& 	 & 0.1 & 
	--- & --- & --- & 115 & 155 & --- & --- & --- & 115 & 160 \\
\hline
1500 GeV & $\gamma\gamma$ & 0.02 & 
	180 & 290 & 540 & 970 & 1020 & 180 & 280 & 530 & 960 & 1020 \\
	& 	 & 0.1 & 
	--- & 140 & 290 & 750 & 840 & 100 & 150 & 300 & 760 & 830 \\
     	& $e\gamma$   & 0.02 &
	--- & 100 & 240 & 560 & 660 & --- & 130 & 240 & 570 & 660 \\
	& 	 & 0.1 & 
	--- & --- & 120 & 360 & 430 & --- & --- & 130 & 360 & 440 \\     
     	& $e^+e^-$    	 &  0.02 & 
	--- & --- & --- & 280 & 315 & --- & --- & --- & 250 & 300 \\
	& 	 & 0.1 & 
	--- & --- & --- & 160 & 200 & --- & --- & --- & 160 & 200 \\
\end{tabular}
\end{center}
\end{table}

\newpage
\begin{figure}
\begin{center}
\centerline{\epsfig{file=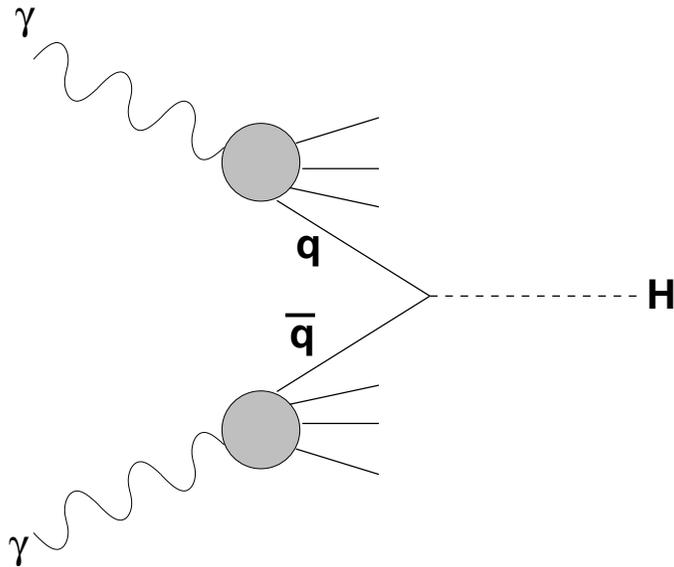,width=3.5in}}
\end{center}
\vspace{20pt}
\caption{Generic Feynman diagram for Higgs boson production via 
the resolved photon process $\gamma\gamma \to q\bar{q}+X \to H +X$.}
\label{Fig1}
\end{figure}

\newpage
\begin{figure}
\begin{center}
\centerline{\epsfig{file=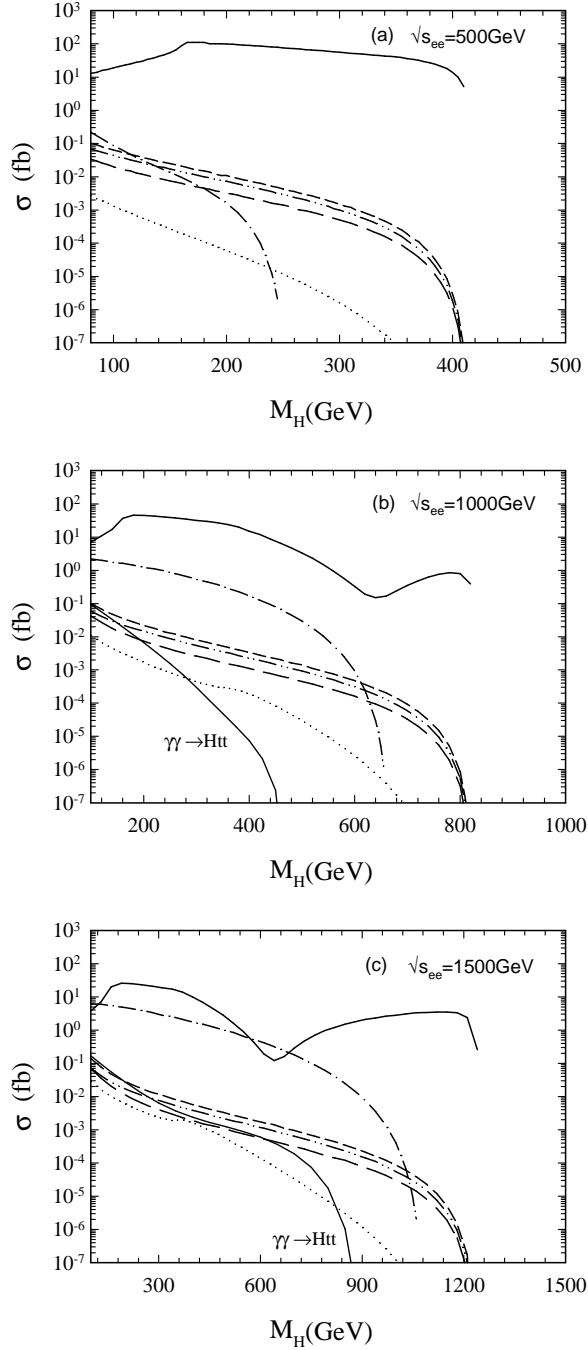,width=3.1in}}
\end{center}
\vspace{20pt}
\caption{Production cross sections for SM Higgs boson for 
(a) $\sqrt{s_{ee}}=500$~GeV, (b) 1~TeV, (c) 1.5~TeV
 with backscattered laser spectrum.  The solid 
line is for $\gamma\gamma\to h$, the short dashed line for 
$\hat{\sigma}(b\bar{b}\to h) +\hat{\sigma}(c\bar{c}\to h)$,
the dot-dot-dashed line for $\hat{\sigma}(c\bar{c}\to h)$, 
the long-dashed line for $\hat{\sigma}(b\bar{b}\to h)$,
the dotted line for $\hat{\sigma}(gg\to h)$ 
and the dot-dashed line for $\hat{\sigma}(WW\to H)$.
$\sigma(\gamma\gamma \to H t\bar{t})$ is shown by the labelled solid 
line.
}
\label{Fig2}
\end{figure}

\newpage
\begin{figure}
\begin{center}
\centerline{
\epsfig{file=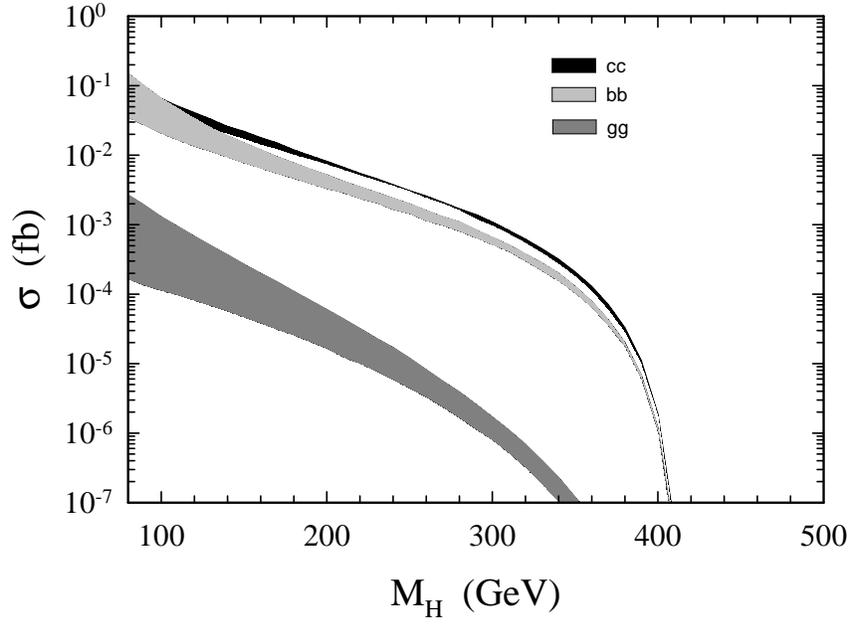,width=4.5in,clip=}
}
\end{center}
\vspace{20pt}
\caption{Variation of the $\gamma\gamma\to H+X$
cross section using different parton distributions for 
$\sqrt{s_{ee}}=500$~GeV with backscattered laser spectrum.  
Details are given in the text.
}
\label{Fig3}
\end{figure}

\newpage
\begin{figure}
\begin{center}
\centerline{\epsfig{file=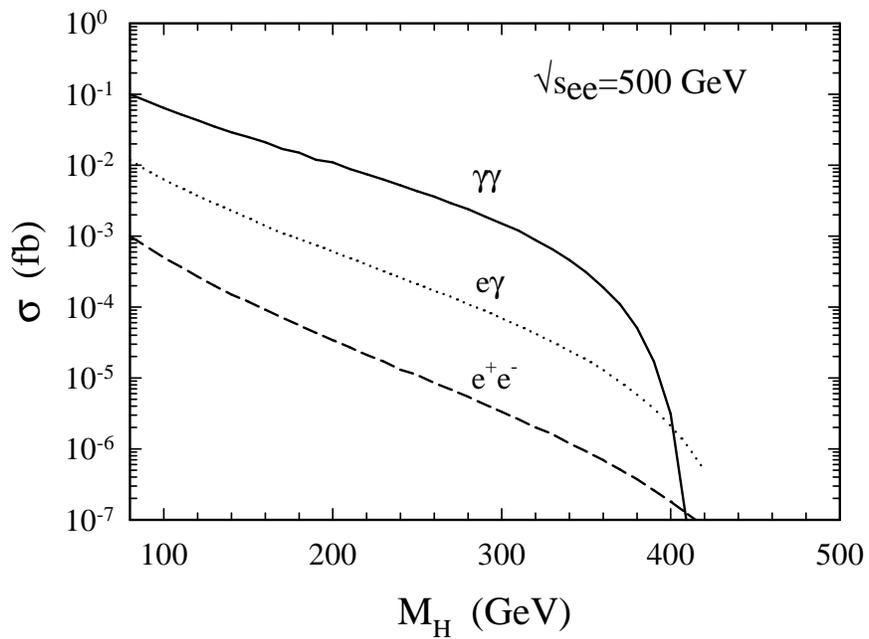,width=4.5in,clip=}}
\end{center}
\vspace{20pt}
\caption{SM Higgs production cross sections via resolved photons with
$\sqrt{s_{ee}}=500$~GeV.  The solid line is for the $\gamma\gamma$ 
case, the dotted line for the $e\gamma$ case, and the dashed line for 
the $e^+e^-$ case.
}
\label{Fig4}
\end{figure}

\newpage
\begin{figure}
\begin{center}
\centerline{\epsfig{file=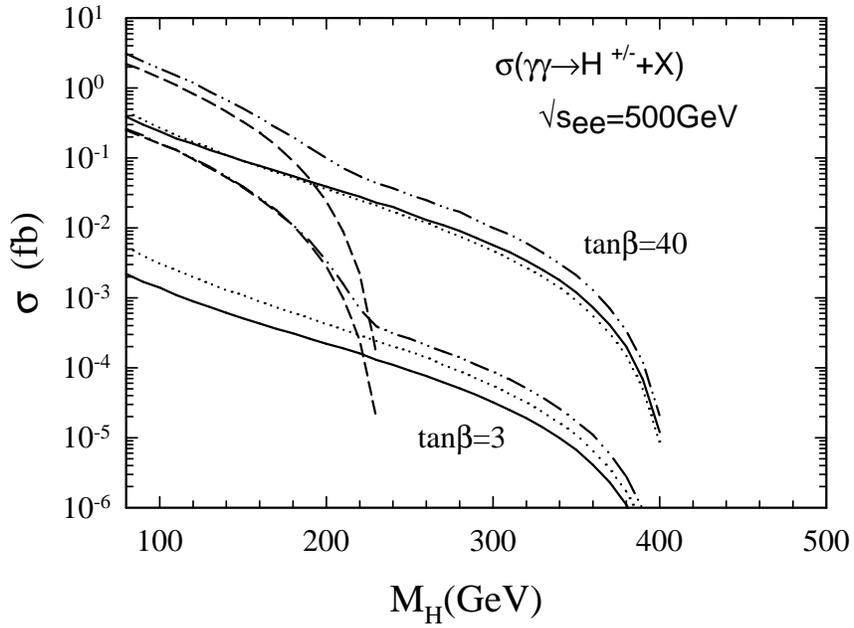,width=4.5in}}
\end{center}
\vspace{20pt}
\caption{Charged Higgs boson production via resolved photon 
contributions in $\gamma\gamma$ collisions at $\sqrt{s_{ee}}=500$~GeV
for $\tan\beta=3.0$ and 40.
The solid lines are for $cb$-fusion, the dotted lines for $cs$-fusion,
the dashed lines for the subprocess $\gamma b \rightarrow t H^\pm$ and the dot-dot-dashed lines for the 
sum of all three contributions to charged Higgs production.  
}
\label{Fig5}
\end{figure}

\newpage
\begin{figure}
\begin{center}
\centerline{\epsfig{file=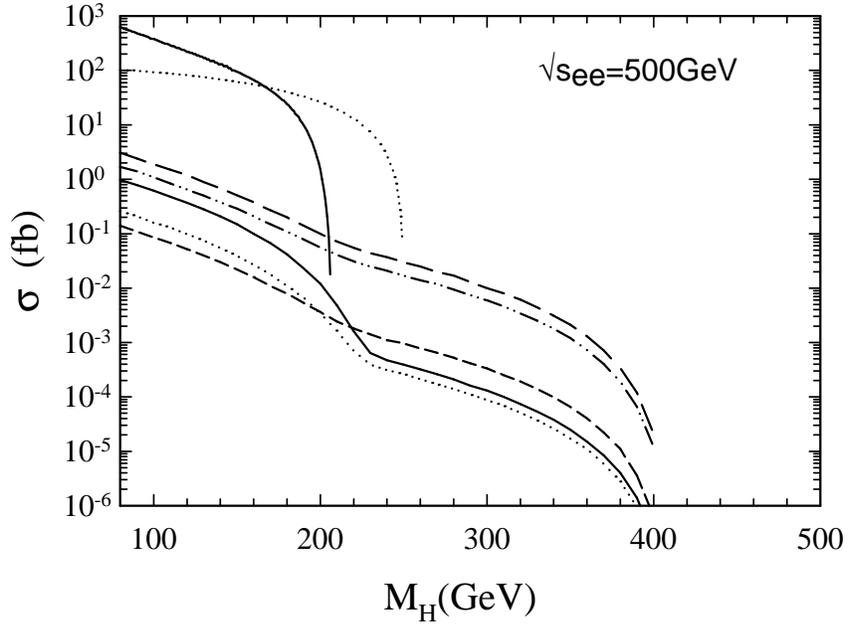,width=4.5in}}
\end{center}
\vspace{20pt}
\caption{Charged Higgs boson production via resolved photon 
contributions in $\gamma\gamma$ collisions at $\sqrt{s_{ee}}=500$~GeV.  
The solid line is for 
$\tan\beta=1.5$, the dotted line for $\tan\beta=3$,
the short-dashed line for $\tan\beta=7$,
the dot-dot-dashed line for $\tan\beta=30$, 
the long-dashed line for $\tan\beta=40$,
the dot-dashed line for $\gamma\gamma\to H^+H^-$,
and the medium dashed line for $e^+ e^-\to H^+H^-$.
}
\label{Fig6}
\end{figure}

\newpage
\begin{figure}
\begin{center}
\centerline{\epsfig{file=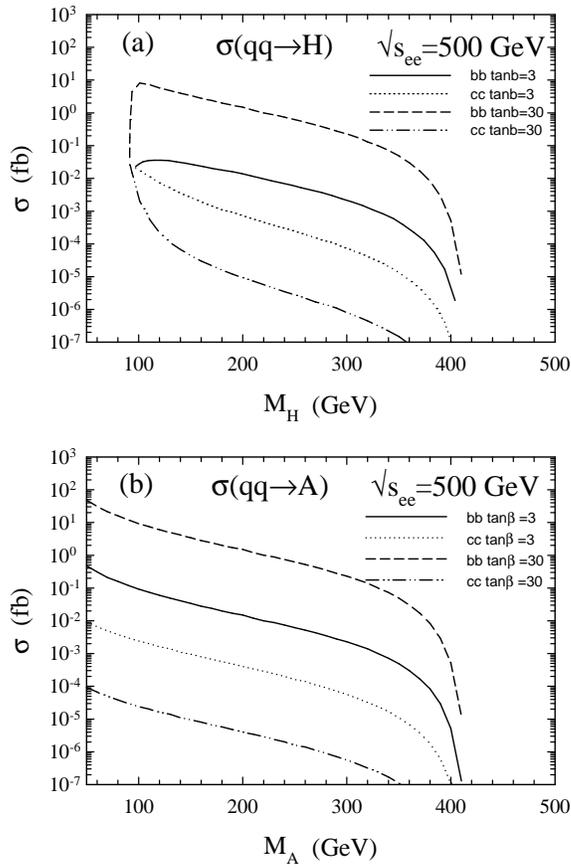,width=4.5in}}
\end{center}
\vspace{20pt}
\caption{Heavy MSSM Higgs boson production cross sections 
via resolved photons for $\sqrt{s_{ee}}=500$~GeV
with backscattered photons.    
(a) is for $\hat{\sigma}(q\bar{q}\to H)$
and (b) is for $\hat{\sigma}(q\bar{q}\to A)$
In both cases
the solid line is for $\hat{\sigma}(b\bar{b}\to H(A))$ for $\tan\beta=3$,
the dotted line is for $\hat{\sigma}(c\bar{c}\to H(A))$ for $\tan\beta=3$,
the dashed line is for $\hat{\sigma}(b\bar{b}\to H(A))$ for $\tan\beta=30$,
and the dot-dot-dashed line is for 
$\hat{\sigma}(c\bar{c}\to H(A))$ for $\tan\beta=30$.
}
\label{Fig7}
\end{figure}

\newpage
\begin{figure}
\begin{center}
\centerline{\epsfig{file=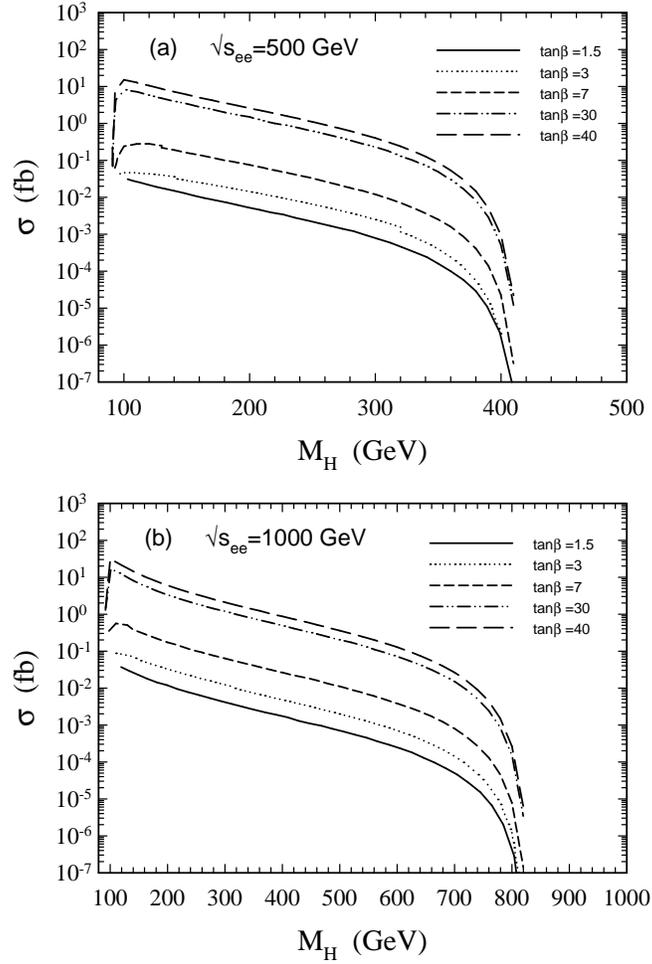,width=4.5in}}
\end{center}
\vspace{20pt}
\caption{
H production cross sections via resolved photons for
$\sqrt{s_{ee}}=500$~GeV with backscattered photons from
$\hat{\sigma}(c\bar{c}\to H) +\hat{\sigma}(b\bar{b}\to H) $.  
The solid line is for $\tan\beta=1.5$, the dotted line for 
$\tan\beta=3$, the short-dashed line for $\tan\beta=7$,
the dot-dot-dashed line for $\tan\beta=30$, and the long-dashed line
for $\tan\beta=40$.
}
\label{Fig8}
\end{figure}

\newpage
\begin{figure}
\centerline{\epsfig{file=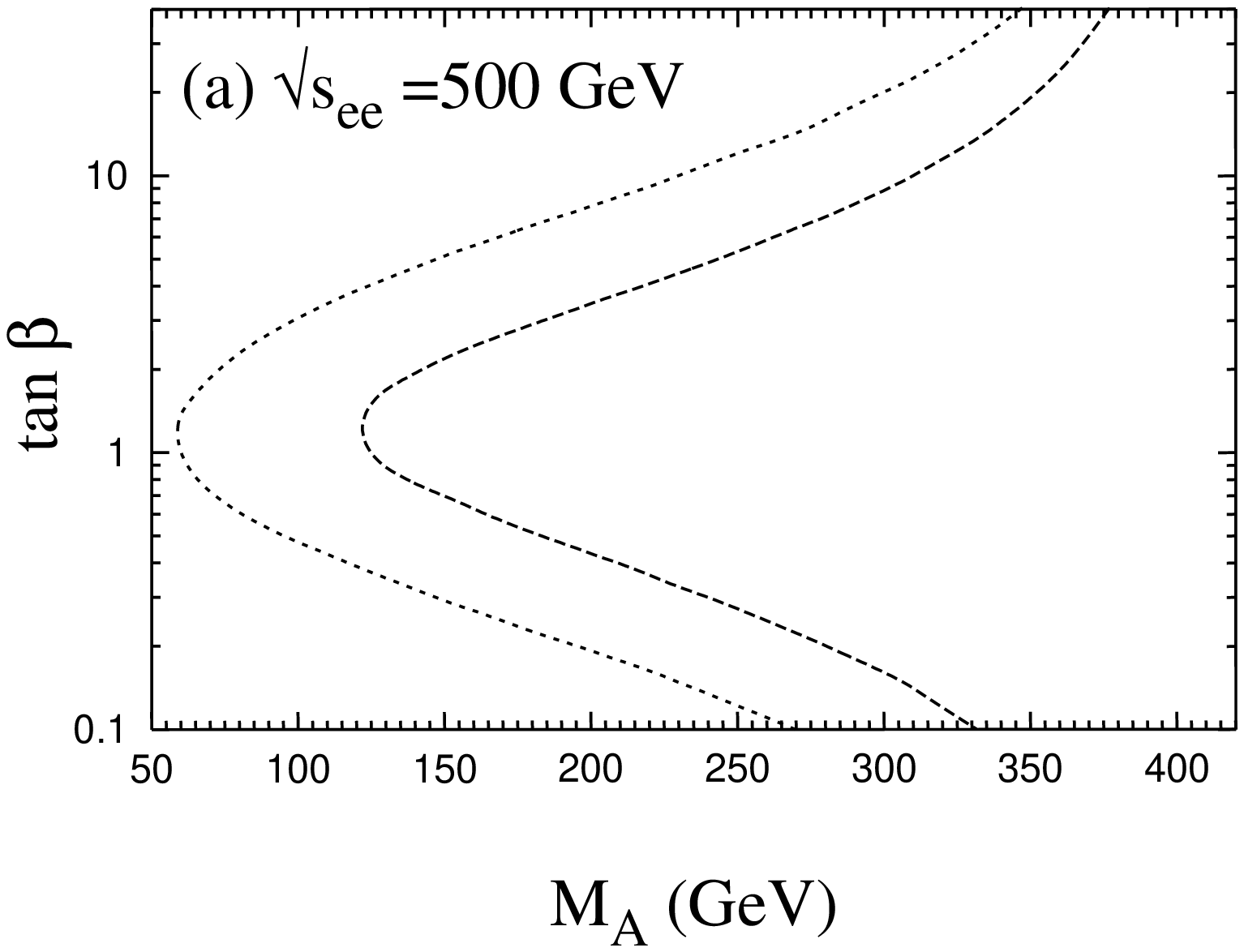,width=4.0in,angle=-90}}
\vspace{10pt}
\centerline{\epsfig{file=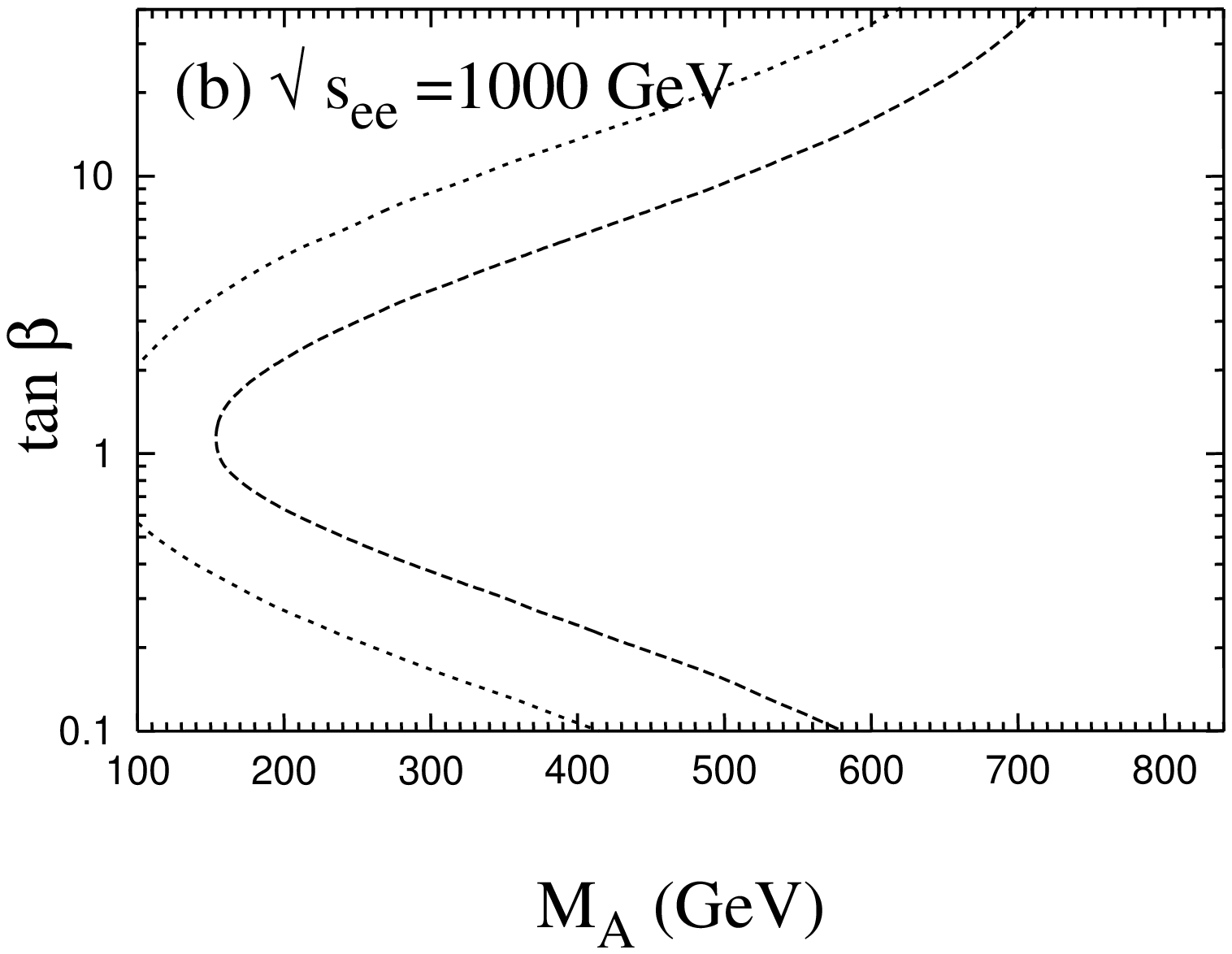,width=4.0in,angle=-90}}
\vspace{20pt}
\caption{
Regions of sensitivity in $\tan\beta -M_A$ parameter space to $A$ 
production via resolved photons with backscattered laser photons. 
The region to the left is accessable and the region to the right is 
unaccessable. 
The dotted line gives  $\sigma = 0.1$~fb contour so that at least 
100~events would be produced in the region to the left 
for 1~ab$^{-1}$ integrated luminosity.
Similary the dashed line 
gives the $\sigma = 0.02$~fb contour  designating the 
the boundary for producing at least $20$~events. 
(a) is for $\sqrt{s_{ee}}=500$~GeV  and (b) for $\sqrt{s_{ee}}=1000$~GeV.
}
\label{Fig9}
\end{figure}


\begin{references}

\bibitem{dawson99}
For a recent review see S. Dawson, 
ICTP Summer School in High-Energy Physics and Cosmology, 
Miramare, Trieste, Italy, 29 Jun -17 Jul 1998. 
[hep-ph/9901280].

\bibitem{carena02}
M. Carena and H.E. Haber, hep-ph/0208209.

\bibitem{susy}
For reviews of the MSSM see
P. Fayet and S. Ferrara, Phys. Rept. {\bf 32}, 249 (1977);
H. P. Nilles, Phys. Rept. {\bf 110}, 1 (1984);
H.E. Haber and G. Kane, Phys. Rept. {\bf 117}, 75 (1985);
M. Drees, hep-ph/9611409;
X. Tata, hep-ph/9807526.

\bibitem{lep2}
ALEPH, DELPHI, L3 and OPAL Collaborations, The LEP
working group for Higgs boson searches, LHWG Note 2002-01 (July 2002);
%
T. Junk, Talk at the 5th International Symposium on Radiative 
Corrections, Carmel CA, Sept 11-15, 2000  hep-ex/0101015.

\bibitem{tevatron}
D0 Collaboration (B. Abbott et al.), hep-ex/9902028;
CDF Collaboration (F. Abe et al.), Phys. Rev. Lett. {\bf 79}, 357 (1997).

\bibitem{run2}
Report of the
Higgs working group of the Physics at Run II Supersymmetry/Higgs
workshop,  hep-ph/0010338.  See also reference 2.

\bibitem{atlas}
ATLAS Collaboration, ATLAS Detector and Physics Performance Technical 
Design Report. CERN-LHCC 99-14.

\bibitem{cms} 
CMS Collaboration. CMS Technical proposal, CERN-LHCC 94-38.

\bibitem{teslatdr}
TESLA Technical Design Report, Part III: Physics at an $e^+e^-$ Linear 
Collider,  Ed. R.D. Heuer, D. Miller, F. Richard, P.M. Zerwas (March 
2001).

\bibitem{higgs}
A recent review of Higgs boson studies is given by
M. Battaglia and K. Desch, hep-ph/0101165.

\bibitem{telnov}
V. Telnov,  Int. J. Mod. Phys. {\bf A13}, 2399 (1998) hep-ex/9802003.

\bibitem{ggtoh}
J.I. Illana LC-TH-2000-002 ({\tt http://www.desy.de/\~{ }lcnotes/});
J.F. Gunion and H.E. Haber, 1990 DPF Summer Study on HIgh Energy 
Physics, Snowmass CO;
D.L. Borden, D.A. Bauer and D.O. Caldwell, Phys. Rev. {\bf D48}, 4018 
(1993);
M. Kr\"amer {it et al.}, Z. Phys. {\bf C64}, 21 (1994)
G. Jikia and S. S\"older-Rembold, Nucl. Phys. Proc. Suppl. {\bf 82}, 
373 (2000);
I.F. Ginzburg, M. Krawczyk, P. Osland, hep-ph/0101208.

\bibitem{ggtomssm}
M. M\"uhlleitner, M. Kramer, M. Spira, P.M. Zerwas, 
Phys. Lett. {\bf B508}, 311 (2001) [hep-ph/0101083];
M.M. M\"uhlleitner, hep-ph/0101177.

\bibitem{melles}
M. Melles, W.J. Stirling, V.A. Khoze, Phys. Rev. {\bf D61}, 054015 (2000).

\bibitem{guo99}
J.-Y. Guo, Y. Liao, and Y-P. Kuang, hep-ph/9912277.

\bibitem{asner}
D.M. Asner, J.B. Gronberg, and J.F. Gunion, hep-ph/0110320.

\bibitem{krawczyk}
P. Niezurawski, A.F. Zarnecki, and M. Krawczyk, hep-ph/0208234.

\bibitem{fei01}
Z. Fei, M. Wen-Gan, J. Yi, L. Xue-Qian, W. Lang-Hui, 
Phys. Rev. {\bf D64}, 055005 (2001).

\bibitem{lqopal}
OPAL Collaboration, (Stefan Soldner-Rembold for the collaboration),
Presented at International Conference on the Structure and the 
Interactions of the Photon (Photon 97) including the 11th
International Workshop on Photon-Photon Collisions, 
Egmond aan Zee, Netherlands, 10-15 May 1997.  hep-ex/9706003.

\bibitem{lqdelphi}
DELPHI Collaboration (P. Abreu et al.),
Phys.Lett. {\bf B446}, 62 (1999).

\bibitem{lq}
M.A. Doncheski, and S. Godfrey,
Phys.Rev. {\bf D49}, 6220 (1994);
Phys.Rev. {\bf D51}, 1040 (1995);
Phys.Lett., {\bf B393}, 355 (1997);
Mod. Phys. Lett. {\bf A12}, 1719 (1997).

\bibitem{fph}
Two recent reviews are: 
R. Nisius, Phys. Rept. {\bf 332}, 165 (2000);
M. Krawczyk, A. Zembrzuski, and M. Staszel, hep-ph/0011083.

\bibitem{backlaser} 
I.F.\ Ginzburg, {\it et al.}, Nucl.\ Instrum.\ Methods, {\bf 205}, 47
(1983); {\it ibid} {\bf 219}, 5 (1984);
V.I. Telnov, Nucl.\ Instrum.\ Methods, {\bf A294}, 72 (1990);
C.\ Akerlof, Report No.\ UM-HE-81-59 (1981; unpublished).

\bibitem{WW}
C. Weizs\"acker, Z. Phys. {\bf 88}, 612 (1934);
E.J. Williams, Phys. Rev. {\bf 45}, 729 (1934).

\bibitem{higgsbr}
A. Djouadi, M. Spira, P.M. Zerwas, Z. Phys. {\bf C70}, 427 (1996);
Phys. Lett. {\bf B311}, 255 (1993);
M. Battaglia, hep-ph/9910271.

\bibitem{mssmdecays}
J. Kalinowski, P.M. Zerwas, Z. Phys. {\bf C70}, 435 (1996).

\bibitem{pdg}
Particle Data Group, 
K. Hagiwara et al, Phys. Rev. {\bf D66}, 010001 (2002).

\bibitem{loops}
L. Resnick, M.K. Sundaresan, and P.J.S. Watson, Phys. Rev. {\bf D8}, 
172 (1973);
M.K. Gaillard and D.V. Nanapouls, Nucl. Phys. {\bf B106}, 292 (1976).

\bibitem{decoupling}
H. Haber, hep-ph/9505240; hep-ph/9707213;
A. Djouadi, V. Driesen, W. Hollik, and J.I. Illana, Eur. Phys. J. 
{\bf C1}, 149 (1998).

\bibitem{grv}
M. Gl\"uck, E. Reya and A. Vogt, {\it Phys. Lett.} {\bf B222}, 149 (1989);
{\it Phys. Rev.} {\bf D45}, 3986 (1992).

\bibitem{cheung93}
K. Cheung, Phys. Rev. {\bf D47}, 3750 (1993).

\bibitem{comphep}
P. A. Baikov et al., Physical Results by means of CompHEP, in Proc. of X
Workshop on High Energy Physics and Quantum Field Theory (QFTHEP-95), eds.
B. Levtchenko, V. Savrin, Moscow, 1996, p. 101 , hep-ph/9701412;
E. E. Boos, M. N. Dubinin, V. A. Ilyin, A. E. Pukhov,
 V. I. Savrin, hep-ph/9503280.

\bibitem{HHguide}
J.~F.~Gunion, H.~E.~Haber, G.~L.~Kane and S.~Dawson,
{\sl The Higgs Hunter's Guide}, 
{\it  Addison-Wesley, Redwood City, USA,  1990}.

\bibitem{barger}
V.~Barger and R.~J.~Phillips,
{\sl Collider Physics},
{\it  Addison-Wesley, Redwood City, USA,  1987}.

\bibitem{cheung94}
K. Cheung, Phys. Rev. {\bf D50}, 4290 (1994).

\bibitem{jikia95}
G. Jikia, Nucl. Phys. {\bf B437}, 520 (1995).

\bibitem{do}
D.W. Duke and J.F. Owens, {\it Phys. Rev.} {\bf D26}, 1600 (1982).

\bibitem{dg}
M. Drees and K. Grassie, {\it Z. Phys.} {\bf C28}, 451 (1985).

\bibitem{lac}
H. Abramowicz, K. Charchula, and A. Levy, {\it Phys. Lett.} {\bf B269}, 458
(1991).

\bibitem{resphot}
M. Drees and R. Godbole, {\it Nucl. Phys.} {\bf B339}, 355 (1990);
M. Gl\"uck, E. Reya and A. Vogt, {\it Phys. Rev.} {\bf D46}, 1973 (1992);
G. A. Schuler, T. Sj\"ostrand, Z. Phys. {\bf C68}, 607 (1995);
Phys. Lett. B {\bf 376}, 193 (1996).

\bibitem{cornet02}
F. Cornet, P. Jankowski, M. Krawczyk, A. Lorca, hep-ph/0212160.

\bibitem{chargedhiggs}
J.F. Gunion, {\it et al.}, Phys. Rev. {\bf D38}, 3444 (1988);
A. Djouadi, J. Kalinowski, P.M. Zerwas, Z. Phys. {\bf C57}, 569 (1993);
A. Djouadi, J. Kalinowski, P. Ohmann, P.M. Zerwas, 
Z. Phys. {\bf C74}, 93 (1997).

\bibitem{komamiya88}
S. Komamiya, Phys. Rev. {\bf D38}, 2158 (1988).

\bibitem{chao93}
D. Bowser-Chao, K. Cheung, and  S. Thomas, 
Phys. Lett. {\bf B315}, 399 (1993) [hep-ph/9304290].

\bibitem{kanemura}
S. Kanemura, S. Moretti, and K. Odagiri, Presented at the Linear 
Collider Workshop 2000, October 24-28 2000, Fermilab USA,
hep-ph/0101354; J. High Energy Phys. {\bf 02}, 011 (2001) 
[hep-ph/0012030].

\bibitem{moretti02}
S. Moretti, and S. Kanemura, hep-ph/0211055.

\bibitem{he02}
H.-J. He, S. Kanemura, and C.-P. Yuan, hep-ph/0209376;
Phys. Rev. Lett. {\bf 89}, 101803 (2002).

\bibitem{doncheski}
M.A. Doncheski, and S. Godfrey, in preparation.

\bibitem{blundell}
H.G. Blundell, S. Godfrey, G. Hay, E.S. Swanson, 
Phys. Rev. {\bf C61}, 025203 (2000).

\bibitem{spira}
M. Spira, Fortschr. Phys. {\bf 46}, 203 (1998).

\bibitem{guo99b}
See also Ref. \cite{guo99}.

\bibitem{mssm}
J.F. Gunion and H. Haber, Nucl. Phys. {\bf B272}, 1 (1986); erratum 
{\bf B402}, 567 (1993); 
Nucl. Phys. {\bf B279}, 449 (1986).

\bibitem{pdb} 
C.\ Caso, {\it et al.}, Particle Data Group, 
Eur.\ Phys.\ J.\ {\bf C3}, 1 (1998).

\bibitem{barger01}
V. Barger, T. Han, J. Jiang, Phys. Rev. {\bf D63}, 075002 (2001).


\end{references}
\end{document}